\documentstyle[11pt,aaspp4,epsfig]{article}

\begin{document}

\title{Evidence for Doppler-Shifted Iron Emission Lines in Black Hole 
Candidate 4U~1630-47}
\author{Wei Cui\altaffilmark{1}, Wan Chen\altaffilmark{2,3}, and 
Shuang~Nan~Zhang\altaffilmark{4}}
\altaffiltext{1}{Room 37-571, Center for Space Research, Massachusetts
Institute of Technology, Cambridge, MA 02139; cui@space.mit.edu}
\altaffiltext{2}{NASA/Goddard Space Flight Center, Code 661,
Greenbelt, MD 20771; chen@milkyway.gsfc.nasa.gov}
\altaffiltext{3}{also Department of Astronomy, University of Maryland, 
College Park, MD 20742}
\altaffiltext{4}{Physics Department, OB 428, University of Alabama in 
Huntsville, Huntsville, AL 35899; Shuang.Zhang@msfc.nasa.gov}

\begin{abstract}
We report the first detection of a pair of correlated emission lines 
in the X-ray spectrum of black hole candidate 4U 1630-47 during its 
1996 outburst, based on RXTE observations of the source. At the peak 
plateau of the outburst, the emission lines are detected, centered 
mostly at $\sim$5.7 keV and $\sim$7.7 keV, respectively, while the 
line energies exhibit random variability $\sim$5\%. Interestingly, 
the lines move in a concerted manner to 
keep their separation roughly constant. The lines also vary greatly 
in strength, but with the lower-energy line always much stronger than 
the higher-energy one. The measured equivalent width ranges from 
$\sim$50 eV to $\sim$270 eV for the former, and from insignificant 
detection to $\sim$140 eV for the latter; the two are reasonably
correlated. 

The correlation between the lines implies a causal connection ---
perhaps they share a common origin. Both lines may arise from a single 
$K_{\alpha}$ line of highly ionized iron that is Doppler-shifted
either in a Keplerian accretion disk or in a bi-polar outflow or even 
both. In both scenarios, a change in the line energy might simply 
reflect a change in the ionization state of line-emitting matter. We 
discuss the implication of the results and also raise some questions 
about such interpretations.
\end{abstract}


\section{Introduction}
Discovered by the Uhuru mission (Jones et al. 1976), 4U 1630-47 is a
well-studied black hole candidate (BHC; Tanaka \& Lewin 1995 and
references therein; also see, e.g., Parmar et al. 1995, 1997, Kuulkers 
et al. 1997, and Oosterbroek et al. 1998 for more recent results). 
The source lies in the general direction of the Galactic center. 
Observations show that the source is heavily absorbed in soft X-rays,
indicating a large distance ($\gtrsim$10 kpc) within the disk of our 
Galaxy. No optical counterpart has yet
been identified, perhaps due to expected large visual extinction ($>$
20 magnitude), so the dynamical evidence for the presence of a black 
hole in the binary system is still missing. 4U 1630-47 is considered 
a BHC only because of its similarities in the X-ray properties to some 
of the dynamically-determined black hole systems (Parmar et
al. 1986). It is a transient X-ray source, like most known BHCs, but 
is one of few such sources that exhibit frequent X-ray outbursts (e.g.
Jones et al. 1976; Priedhorsky 1986; Kuulkers et al. 1997). 

The canonical X-ray continuum of a BHC consists of two components (see
review by Tanaka \& Lewin 1995 and references therein): an ultra-soft 
component at low energies ($<$ 10 keV) and a power-law component 
at high energies, with photon indices roughly in the range of
1.5--2.5. The shape of the former is roughly that of a blackbody with 
temperature a few tenths to $\sim$2 keV, and has been successfully 
modeled by X-ray emission from the innermost portion of
an optically thick (but geometrically thin) accretion disk surrounding 
the central black hole, while the latter is generally attributed to 
inverse-Comptonization processes due to the presence of energetic 
electrons (thermal or non-thermal or both) in the binary system. For 
transient BHCs, the disk component dominates the overall X-ray energy 
output over the Compton component near the peak of an X-ray outburst
(which is similar to the soft state for persistent BHCs like Cyg X-1, 
in terms of the observed X-ray properties). Observations show that the 
power-law component is steep (with photon index $\sim$2.5) and extends 
to nearly 1 MeV without any apparent breaks
(Grove et al. 1998), which would seem to signify the role of
non-thermal Comptonization in the X-ray production processes (Coppi 
1999, and references therein). As the 
outburst decays, the power-law component becomes flatter and thus 
increasingly dominant in the energy distribution; it also breaks at
tens to several hundred keV, which is the signature of thermal 
Comptonization. The studies of 4U 1630-47 have shown that, for the 
most part, the source fits this description.

Emission lines, as well as absorption lines and edges, are sometimes
observed in the X-ray spectrum of BHCs (e.g., Barr et al. 1985; 
Kitamoto et al. 1990; Done et al. 1992; 
Ebisawa et al. 1996; Cui et al. 1997, 1998; Ueda et al. 1998). These
narrow spectral features appear in the energy range 6--8 keV and are 
usually attributed to emission or absorption processes involving iron
K-shell electrons. Although 
the exact location of line-emitting matter is often debatable, there 
is evidence, at least for some BHCs (e.g., Fabian et al. 1989; Cui et 
al. 1998; \.{Z}ycki et al. 1999), that the 
observed iron $K\alpha$ line originates in the innermost part of the 
accretion disk, close to the black hole. If this proves to be the case, 
the profile of such lines would be distorted by the strong gravitational 
field of the hole (e.g., Fabian et al. 1989) and could then be carefully 
modeled to possibly derive the intrinsic properties of the hole, such 
as its angular momentum (Laor 1991; Bromley et al. 1997). Also observed 
are Doppler-shifted emission
lines from the relativistic jets of SS 433 (Kotani et al. 1996), which 
is sometimes thought to contain a neutron star, although this is highly
debatable (Mirabel \& Rodriguez 1999). The studies of such lines have 
proven fruitful in gaining insights into the physical properties of
the jets (Kotani et al. 1996). The emission lines could, therefore,
become a valuable tool for probing the immediate surroundings of X-ray 
emitting regions, in terms of its geometry and kinematics, and provide
important diagnostics of matter there, such as its abundance and
ionization state.
In this paper, we report, for the first time, the detection of a pair
of correlated emission lines in the X-ray spectrum of 4U 1630-47
during its 1996 outburst, using archival data from the Rossi X-ray
Timing Explorer (RXTE). 

\section{Observations}
\begin{figure}
\psfig{figure=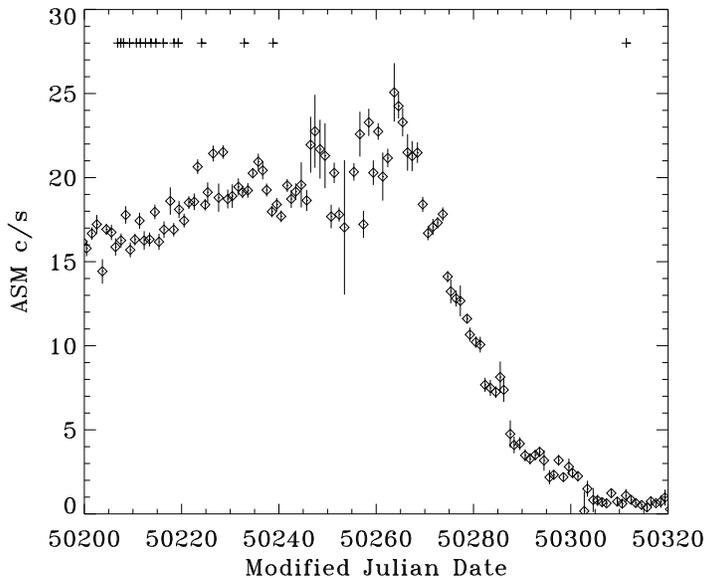,width=4in}
\caption{ASM light curve of 4U 1630-47 during its 1996 X-ray outburst. Each 
data point represents an one-day averaged measurement. As a reference, the Crab 
produces about 75 c/s.The crosses at the top indicate the times of the 
pointed RXTE observations.}
\label{fg:asm}
\end{figure}

Fig.~\ref{fg:asm} shows the light curve of 4U 1630-47, obtained by 
the All-Sky Monitor (ASM) aboard RXTE, that highlights the portion of 
its ``flat-topped'' 1996 X-ray outburst when the source was also 
observed by RXTE's main instruments.\footnote{The full ASM 
light curve can be obtained from the public archival database 
maintained by the RXTE Guest Observer Facility through their web site 
at http://heasarc.gsfc.nasa.gov/docs/xte/asm\_products.html.} The 16
pointed RXTE observations (as marked in Fig.~\ref{fg:asm}) cover a 
good portion of the peak plateau of the outburst, initially at a pace 
of once per day but the pace is much reduced towards the end of the 
monitoring campaign. The last observation occurred near the end of 
the decaying phase, about two months after the onset of the outburst. 
This particular observation has provided a valuable dataset for 
studying the transition period as the source was on its way back to 
the quiescent state and for comparing the X-ray properties between a 
``low state''and a ``high state'' (i.e., during peak of the outburst).

RXTE carries two pointing instruments, the Proportional Counter Array
(PCA) and the High-Energy X-ray Timing Experiment (HEXTE). The PCA 
covers a nominal energy range 2--60 keV and the HEXTE 15--250 keV. 
Although our main focus in this study is on spectral lines, we have 
examined data from both instruments, hoping to better constrain the 
underlying continuum with a wider energy band. Unlike typical BHCs, 
however, the observed X-ray spectrum of 4U 1630-47 at the peak of the 
outburst is so steep (few PCA counts above $\sim$30 keV) that the 
short HEXTE exposures provide little additional information. Hence, 
we present the results from analyzing only the PCA data. 
Table~\ref{tb:log} summarizes the key parameters of the PCA observations. 
Note that the PCA consists of five individual Proportional Counter Units 
(PCUs), but not all PCUs were turned on during all observations. 

\section{Data Analysis and Results}
We performed spectral analysis using the {\it Standard 2} data. For
all PCUs, the calibration of the first xenon layer has achieved the 
most satisfactory results, while that of other two layers is more
uncertain, especially in the energy range where iron spectral features
are often observed. Therefore, we based our analysis on data from the 
first xenon layer only.

We used the calibration files (including response matrices and
background models) delivered with the most recent release of FTOOLS 
(version 4.2). For each observation, we constructed separate spectra 
(both source+background and background only) for the first layer of 
each PCU. We then
carefully examined the quality of background models, under the
assumption that the observed counts in the highest energy channels can
be attributed entirely to the background. We found that the results
were quite satisfactory except for two observations. In those two cases
(observations 3 and 4 in Table~\ref{tb:log}), the background spectrum
seems to be of the correct shape but the overall normalization is a
bit too low. We have tried but failed to find any obvious culprits for
the problem. For these two observations, we simply re-normalized the 
background spectrum (by a factor of 1.08 and 1.15, respectively) so 
that it matches the sum spectrum at the highest energy channels. This
procedure seems to work well in the sense that the quality of
background subtraction achieved for these two observations appears to 
be comparable to that for others. We also note that for observations 
14 and 15, although the background spectrum looks reasonable on average, 
the individual data points scatter much more than usual at high energies.
The cause for such excessive scatter is not clear either.

An important calibration issue is the lack of satisfactory estimation 
of the systematic uncertainties in the response matrices of the PCA. 
A common practice is to add a certain amount (e.g., 1\% of total
counts) in quadrature to Poisson errors uniformly across the entire 
energy range. We know that the uncertainty is significantly larger 
near xenon absorption edges, so this approach does not always work 
well for RXTE observations of bright sources, which are dominated by 
systematic uncertainties. In fact, this leads to a more fundamental 
problem about the applicability of the $\chi^2$ statistics to the 
modeling of such RXTE data. While we can perhaps still use this 
statistics to compare different models to a certain degree, one 
should be cautious
of relying on such statistics alone to reject or accept a certain
model. Unfortunately, we are currently not aware of any more objective 
ways of estimating systematic uncertainties in the calibration of the 
PCA. Since our primary objective here is to study spectral lines, 
instead of modeling the continuum, we have decided to simply follow 
the common practice by adding 1\% systematic error to the data. For 
this study, we are mostly concerned with the instrumental artifacts
that are localized near the xenon edges, because they could mimic 
line-like features. For the first xenon layer, the known artifacts 
seem to be quite small. 
 
\begin{figure}
\psfig{figure=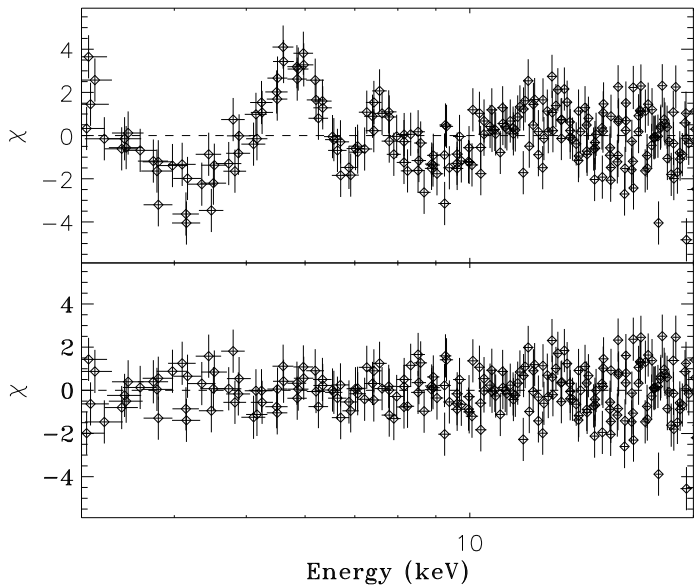,width=4in}
\caption{Non-dip residual plots for Observation 1: (upper panel) the
continuum is modeled by a multi-color disk and a power law, and (lower
panel) the same continuum model plus two Gaussian functions.}
\label{fg:res}
\end{figure}

\begin{figure}
\psfig{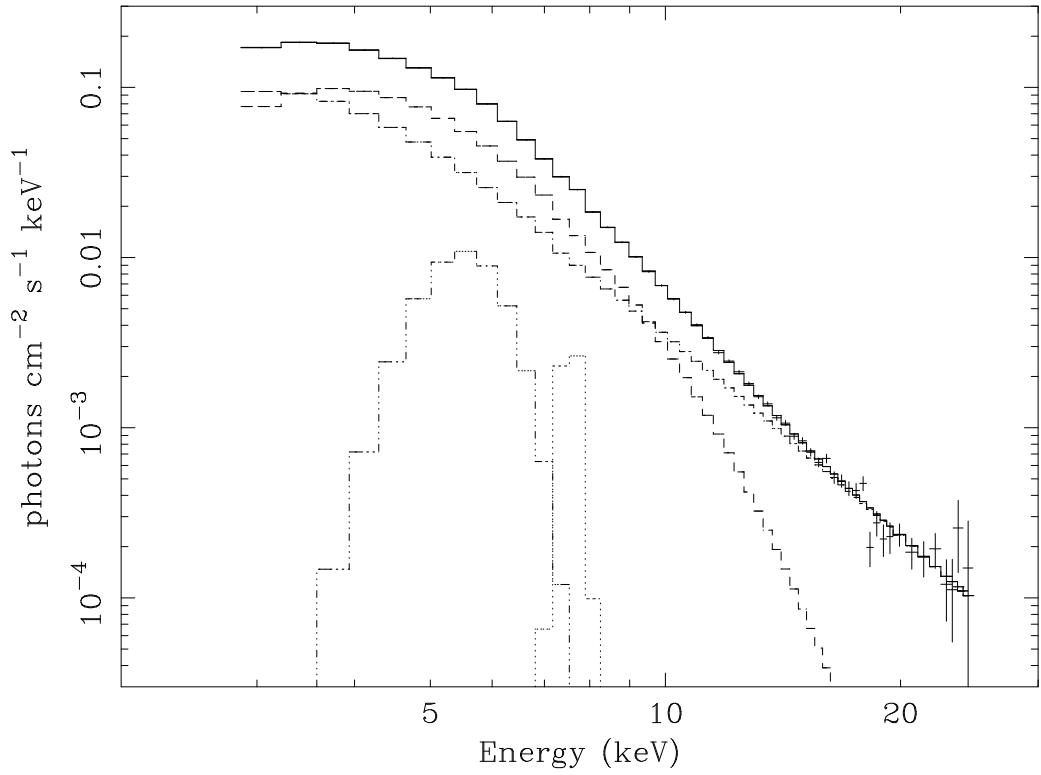}
\caption{Unfolded X-ray spectrum for Observation 1. For clarity, the
spectra and model fits for individual PCUs have been co-added (for 
display only). Also shown is contribution from each model component 
(see text for a description of the model adopted). }
\label{fg:spec}
\end{figure}

After subtracting the background spectrum from the sum, we proceeded
to model the resulted source spectrum. For the continuum, we adopted a
composite model of a multi-color disk (``diskbb'' in XSPEC; e.g.,
Mitsuda et al. 1984) and a
power law, which is typical of BHCs. We then simultaneously applied
the model to the first-layer spectra of all PCUs, with
the relative (to PCU 0) normalization of the PCUs floating to account 
for any slight mismatches among the detectors. All the results are, 
therefore, normalized to the absolute calibration of PCU 0. This model 
alone fails to fit the observed X-ray spectrum in all cases (see the 
upper panel of Fig.~\ref{fg:res} for Observation 1); the residual plot 
reveals two line-like features between 5 and 8 keV. The fit becomes 
much improved (and formally acceptable in terms of the $\chi^2$ 
statistics) after two Gaussian functions have been added to the model 
(see the lower panel of Fig.~\ref{fg:res}). Fig.~\ref{fg:spec} shows 
the unfolded spectrum for Observation 1, along with each individual 
components of the model. The lines are located at $\sim$5.5 keV and 
$\sim$7.5 keV, respectively, with equivalent widths $\sim$188 eV and 
$\sim$76 eV (see Table~\ref{tb:line}). We carried out the same
analyses for other observations. The results are summarized in 
tables~\ref{tb:cont} and \ref{tb:line} for the continuum and lines, 
respectively. 

\begin{figure}
\psfig{figure=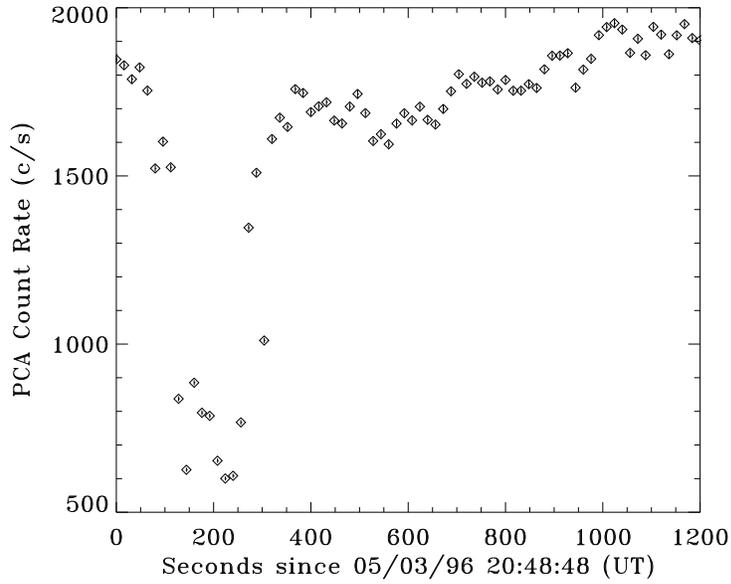,width=4in}
\caption{Light curve for Observation 1. Note an intensity dip during
the early part of the observation.}
\label{fg:dip}
\end{figure}

\begin{figure}
\psfig{figure=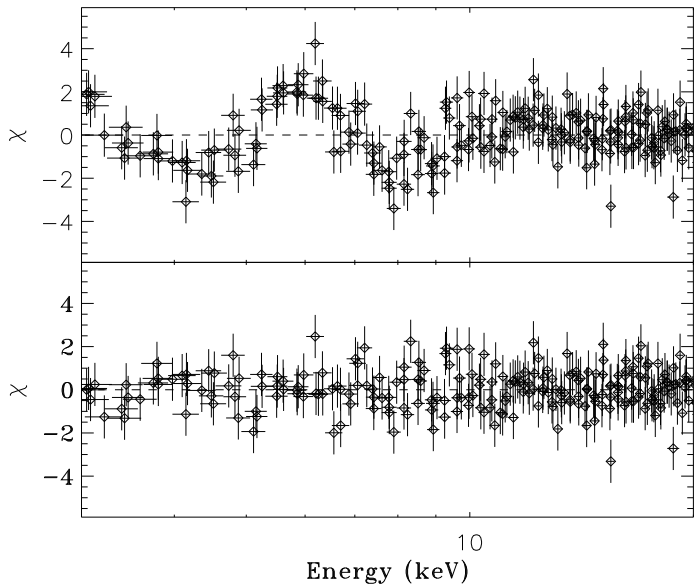,width=4in}
\caption{The same as Figure~2, but for the dip. Note that only the 
lower-energy line is apparent here. }
\label{fg:dres}
\end{figure}

During the first
observation, a large ``dip'' was observed in the X-ray intensity of
the source, as shown in Fig.~\ref{fg:dip}. It has been suggested 
that the dip can be accounted for by a partial covering
of the X-ray emitting disk by some passing-by clouds, and that the 
complicated structures within the dip simply reflect the change in 
the absorbing column density through the clouds (Tomsick et al
1998). Although it is not our primary goal here to study the origin 
of the dip, for the purpose of comparison (the dip was excluded 
from the analysis already described), we constructed an average X-ray 
spectrum for the dip and conducted similar analyses with the same 
continuum model. We found that the dip spectrum can be described 
satisfactorily (in terms of the reduced $\chi^2$ value) with the 
addition of only one Gaussian function (Fig.~\ref{fg:dres}). The 
Gaussian seems to 
correspond to the lower-energy line seen in the non-dip spectrum, 
although the line seems much broader and stronger. The results for 
the dip are summarized in Table~\ref{tb:dip}. In this model, no
significant change is required in the line-of-sight column density 
during the dip (by comparing Table~\ref{tb:dip} to Table~\ref{tb:cont}). 
Instead, {\it only} the disk component seems to have varied 
significantly: the temperature at the inner edge of the disk
becomes higher, but the overall normalization is lower, which perhaps
indicates that the inner edge of the disk moves closer to the central
black hole. This would be opposite to some of the dips observed in 
black hole candidate GRS 1915+105, which have been attributed to a 
viscous instability that causes the disruption of the inner part of 
the disk (Belloni et al. 1997). No similar dips are seen during the 
subsequent RXTE observations of 4U 1630-47.

\begin{figure}
\psfig{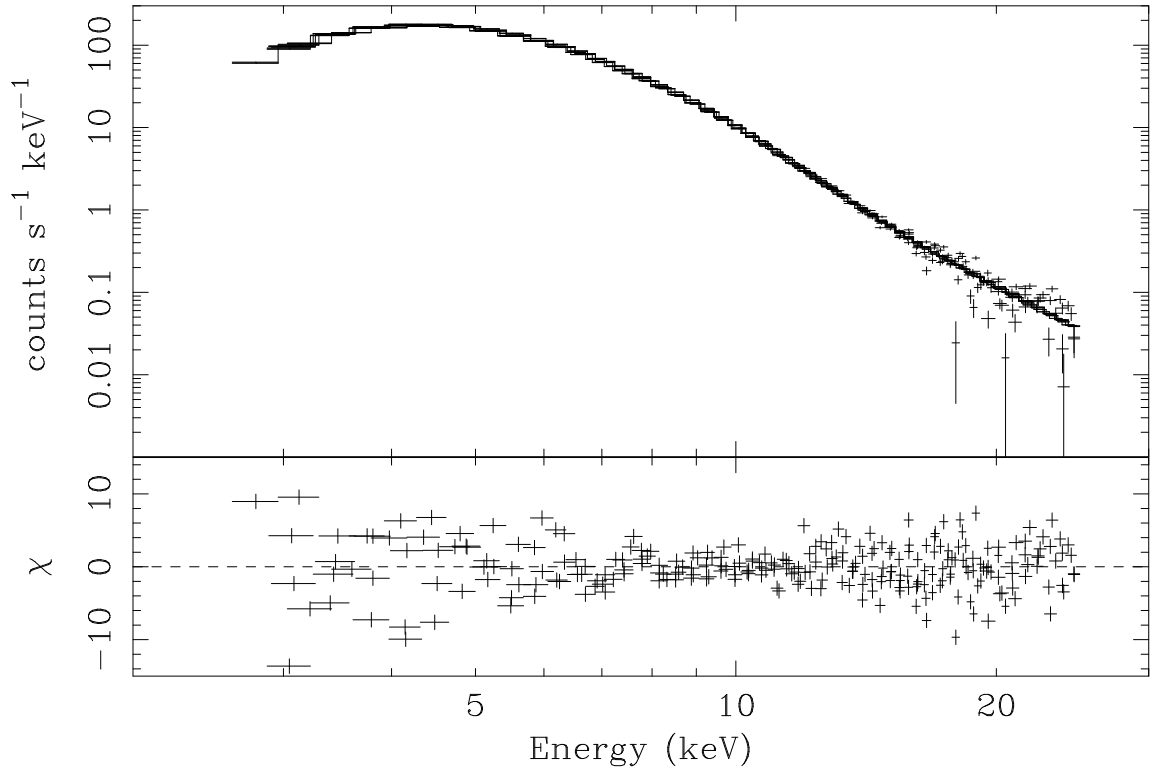}
\caption{Measured X-ray spectrum from Observation 14. The solid
histogram shows the best-fit model, and the residuals of the fit are
shown in the lower panel. Note large residuals at low and high
energies. }
\label{fg:o14}
\end{figure}

It is clear, from tables~\ref{tb:cont} and \ref{tb:line}, that the
model adopted seems to describe the observed X-ray spectra of 
4U 1630-47 
quite well, except for observations 10, 14, and 15. For Observation
10, the large reduced $\chi^2$ are almost entirely due to significant 
structures in the residual at low energies ($<$ 10 keV); Observation
14 suffers from the same problem, in addition to large residuals at
high energies ($>$ 20 keV) which are caused by excessive scatter of
data points in the background spectrum already mentioned; the latter 
also explains the problem for Observation 15. These problems are
illustrated in Fig.~\ref{fg:o14}, taking Observation 14 as an
example. The magnitude of the low-energy 
structures in observations 10 and 14 are quite small, compared to 
the emission lines detected. While the features might represent 
interesting real effects, we will not pursue them any further in this 
study. 

The continuum of 4U 1630-47 is unusually soft during the peak of the 
1996 outburst, with power-law photon 
indices in the range of 3--5 (for comparison, the photon index is 
typically around 2.5 for BHCs). The power-law component varies
significantly during the outburst. The disk component appears quite 
typical of BHCs, with the temperature $\sim$1.3 keV at the inner 
edge of the disk. The high line-of-sight column density may simply 
indicate the location of the source: it is buried in the disk of our 
Galaxy and far away from us, as also concluded from other studies.

\begin{figure}
\psfig{figure=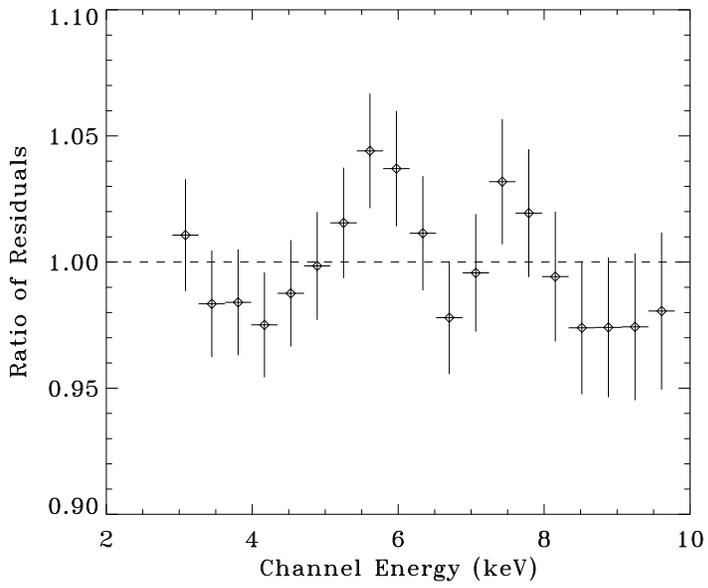,width=4in}
\caption{Ratio of the residual plot of 4U 1630-47 to that of the Crab. 
The residual plot is obtained by taking the ratio of the data
(from the first xenon layer of PCU 0 only) to the model. For 4U
1630-47, the data is from Observation 1 and the model consists of 
a multicolor disk and a power law (with absorption); for the Crab, 
the model is an absorbed power law.}
\label{fg:ratio}
\end{figure}

\begin{figure}
\psfig{figure=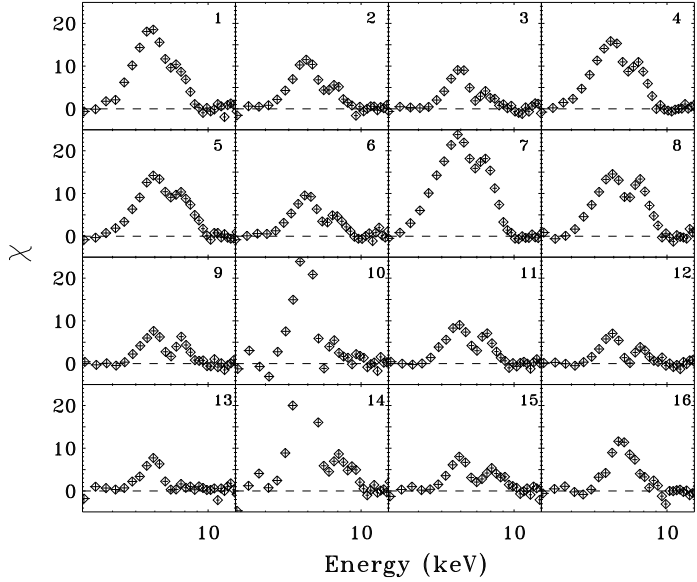,width=4in}
\caption{Detected emission lines. Each panel, labeled by the
observation number, shows a residual plot after the underlying
continuum is subtracted from the source spectrum. As in Fig.~3,
individual PCU spectra have been co-added. Note potential large
systematic uncertainty in observations 10 and 14 (see text). }
\label{fg:lines}
\end{figure}

Although a pair of emission lines are detected in all observations 
during the peak of the outburst (the first 15 in Table~\ref{tb:log}), 
in some cases, the higher-energy line is quite weak and thus the 
detection is not very significant. For such cases, the derived line 
equivalent width can be significantly affected by the calibration 
uncertainties. It is clear, however,
that the detected lines cannot be entirely attributed to
instrumental artifacts, because they vary greatly from observation
to observation and show no apparent correlation with 
the exposure time (which roughly determines the signal-to-noise 
ratio of the data here). To 
further investigate possible artifacts caused by calibration 
uncertainties, we analyzed a PCA observation of the Crab (Observation ID 
10200-01-20-00) which was made during the same epoch (on 01/31/97), 
following exactly the same procedure (e.g.,
only the first xenon layer of each PCU is used). The X-ray spectrum
is well described by an absorbed power law and the
derived model parameters are all within the nominal values for the 
source ($N_H = 3.1\times 10^{21}\mbox{ }cm^{-2}$, photon index $2.16$, 
and normalization $12.5\mbox{ }ph\mbox{ }keV^{-1}\mbox{ }cm^{-2}\mbox
{ }s^{-1}$). The residuals of the fit can be quantified by the ratio
of the data to the model. We obtained such residual plots for both 
the Crab and 4U 1630-47 (from the continuum model without Gaussian
functions), and the ratio between the two plots should roughly be 
free of any calibration uncertainties. One such ratio plot (for 
PCU 0 alone) is shown in Fig.~\ref{fg:ratio}. The two emission 
lines are clearly present. Assuming that the spectrum of the Crab 
is featureless, we conclude that both lines are physically associated 
with 4U 1630-47, even though the inferred line parameters (physical 
width and equivalent width, in particular) may be sensitive to 
calibration uncertainties. Fig.~\ref{fg:lines} highlights the
detected emission lines in all observations after the underlying 
continuum being subtracted from the source spectrum.

Only one emission line is present in the last observation (\#16, as
listed in Table~\ref{tb:log}; see also Fig.~\ref{fg:lines}), as the
source is just about to return to the quiescent state. Interestingly, the
line is located roughly midway between the pair of lines seen during the
peak of the outburst, and it appears to be much stronger. The measured
power-law photon index ($\sim$2) is typical of transitional periods
between ``high'' and ``low'' spectral states in BHCs (e.g., Cui et
al. 1997). The X-ray emitting portion of the disk seems to be much
cooler, and the inner disk edge becomes farther away from the central
black hole. The measured column density is $\sim$30\% lower than that
during the peak. Similar evolution of the column density was also noted 
in another study of 4U 1630-47 during the decaying phase of its 1998 
outburst (Oosterbroek et al. 1998). It was speculated that a
substantial amount of material could be produced at the onset of the 
outburst, which, subsequently, is either accreted onto the black hole 
or expelled from the binary system. 
Alternatively, the larger column density might indicate a significant 
increase in the scale height of the accretion disk during the
outburst, if the binary system is highly inclined with respect to 
the line of sight (as speculated by Kuulkers et al. 1998). 

\section{Discussion}
\begin{figure}
\psfig{figure=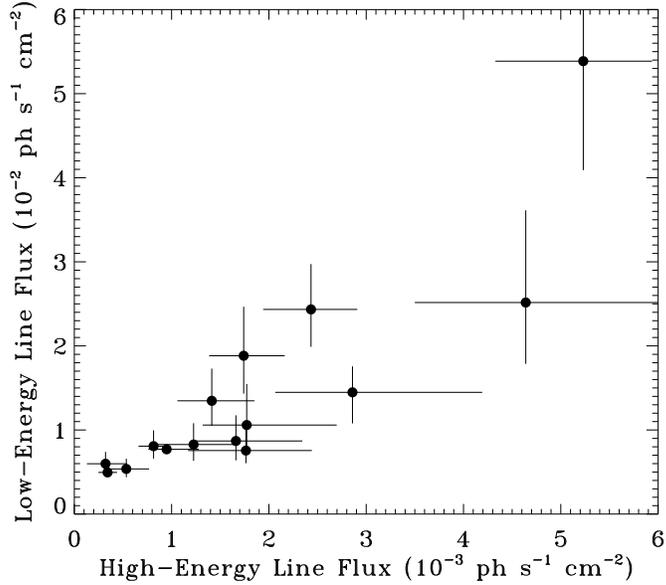,width=4in}
\caption{Correlation between the fluxes of the two emission
lines. Note that the error bars are likely underestimated at low
fluxes, due to remaining systematic uncertainties. }
\label{fg:l1l2}
\end{figure}

The RXTE observations of 4U 1630-47 have revealed the presence of two
emission lines when the source was near the peak of its 1996 X-ray 
outburst. Although the lines seem to move around at random, they 
move in unison, so as to keep their separation roughly 
constant (see Table~\ref{tb:line}). Also, the lines vary 
significantly in strength, but with the lower-energy line always 
much stronger than the higher-energy one. The measured line fluxes 
are reasonably correlated, as shown in Fig.~\ref{fg:l1l2}. 
The correlation between the lines seems to imply a causal connection
--- perhaps they share a common origin. Probably, both  
originates in a single $K_{\alpha}$ line from highly ionized iron 
that is Doppler-shifted either in a Keplerian accretion disk or in a 
bi-polar outflow or even both. The required line-of-sight velocity 
is only roughly
$0.15 c$, which can easily be accommodated in both scenarios. Although 
the detailed line properties (such as profiles) are expected to be 
different for the two cases, the lack of adequate energy resolution 
of the data makes it impossible to favor one over the other at
present. In both cases, a change in the line energy might be due to 
the variation in the ionization state of the line-emitting matter.

If the observed emission line pair is from a single, double-peaked 
iron $K_{\alpha}$ line that originates in the accretion disk close to 
the central black hole, both gravitational and transverse
Doppler shifts would tend to move line photons toward lower energies. 
In this case, a stronger red peak may only reflect the skewness of the 
line profile which is not resolved here, again due to the lack of 
energy resolution. Further support for a disk origin of the line pair 
can be derived from the fact that the higher-energy line appears to 
be much narrower than the lower-energy one in most cases (see 
Table~\ref{tb:line}). This favors a line profile with a much 
more extended red wing (than the blue wing), which is characteristic 
of a disk line. Moreover, during the dip, the inner edge of the disk 
appears to have moved closer to the black hole (see discussion in
\S~3). We would, therefore, expect that the red peak grows stronger, 
due to stronger gravitational redshift; this indeed appears to be the 
case (see Table~\ref{tb:dip}). For a disk line, the highly blue-shifted 
peak would necessarily require a rather high inclination angle of the 
accretion disk (e.g, Fabian et al. 1989), which would be consistent 
with the dipping activity observed of 4U 1630-40 if the dip is caused
by absorption (see arguments by Kuulkers et al. 1998).
The presence of broad (or "smeared") 
iron lines from the accretion disk of other BHCs has previously been 
suggested by \.{Z}ycki and his co-workers in a series of papers 
(e.g., \.{Z}ycki, Done, \& Smith 1999, and references therein), by 
modeling the reflection component of the X-ray spectrum.

The outflow model, on the other hand, cannot naturally explain 
why the red-shifted line is so much stronger than the blue-shifted
one, since the opposite is expected as the result of Doppler
boosting. However, the problem is not fatal, due to our ignorance of
astrophysical jets and the physical processes therein. It is known 
that the measured radio fluxes from the receding and approaching 
jets in ``micro-quasars'' are not always consistent with Doppler 
boosting (Hjellming \& Rupen 1995; Fender et al. 1999). In fact, the 
receding jet sometimes appears brighter than the approaching
one. Sometimes, the relative brightness of twin jets flip-flops as the
jets evolve (see Fig.~2 in Fender et al. 1999). These ``anomalies'' 
almost certainly reflect the intrinsic difference between the two
jets. Similar difference could also have existed in a bi-polar 
outflow from 4U 1630-47, to account for the observed line flux ratio. 
Moreover, the asymmetry in the environment surrounding either side of 
the outflow might cause, for example, the emission from the
approaching flow being obscured more than that from the receding one, 
independent of any other physical processes. Therefore, the
outflow scenario is by no means ruled out, although a disk origin
of the observed emission lines does seem more likely.

Only one emission line is seen during the transition of the source to
the quiescent state. For a line that originates in the innermost 
portion of the accretion disk, it would imply that the inner edge of 
the disk has moved farther away from the black hole (and thus the 
effects of Doppler shift become small) as the transition proceeded,
consistent with the inferred evolution of the disk from the model fit
(see \S~3). It is worth pointing out that such evolution of the 
accretion disk agrees qualitatively with the expectation of the 
``advection-dominated accretion flow'' (ADAF) models for state transitions 
in BHCs (e.g., Esin et al. 1998). If the line originates in a bi-polar 
outflow, on the other hand, a single-peaked line would imply that the 
velocity of the outflow is much reduced during the transition. Given 
the limited spectral resolution of the PCA, it is also possible that 
the observed lines are a mixture of emission lines both from the inner 
region of the accretion disk and from a bi-polar outflow which might 
only occur during the peak of the outburst. In this case, the
evolution of the emission line from being double-peaked to 
single-peaked, as the source approaches the quiescent state, could 
be attributed to the cessation of the outflow and the receding of the 
inner edge of the accretion disk during the process. During the
transition, the increase in the equivalent width of the line (see
Table~\ref{tb:line}) is probably due to the combination of a much 
flatter power-law component and a larger solid angle subtended by a 
larger Comptonizing region (as required, e.g., by the ADAF models).

It is interesting that no emission lines have ever been reported of 4U
1630-47 previously. This can either be attributed to the much improved
spectral capability of the RXTE instrumentation, especially the large
effective area of the PCA which is essential for detecting relatively
weak, broad lines, or to the rare occurrence of such spectral
features. Indeed, the observed X-ray properties of 4U 1630-47 during
the 1996 outburst appear to be somewhat unusual, compared to other
transient BHCs in outburst. Most notably, the observed X-ray continuum 
is extremely soft: hardly any counts are detected above $\sim$30
keV. At high energies, the continuum can still be well described by a
single power law but the photon index reaches as high as $\sim$5,
compared to typical values of $<$ 3 for BHCs. In 1998, 4U 1630-47
experienced another X-ray outburst, which was well covered by the
ASM/RXTE,\footnote{see, again, 
http://heasarc.gsfc.nasa.gov/docs/xte/asm\_products.html for the 
ASM light curve.} as well as by the RXTE's pointing
instruments. We will report the results for this outburst in a future
publication (Cui et al., in preparation). The preliminary spectral 
results from the PCA data give no indication for the presence of a 
similar emission-line pair during the outburst. The observed power-law 
continuum is quite typical of BHCs, which was also noted based on
the results from BeppoSAX observations (Oosterbroek et al. 1998). 
Perhaps, the 1996 outburst is an unusual one for 4U 1630-47, in which 
a bi-polar outflow might indeed have been formed. The source was
observed at radio wavelengths with the {\it Australia Telescope 
Compact Array} and the {\it Very Large Array} during this period, 
but no emission was detected (Kuulkers et al. 1997; Hjellming 
et al. 1999, although radio emission was observed during the 1998 
outburst, which the authors suggested as evidence for the presence 
of jets).

Finally, it seems reasonable to ask why no Doppler-shifted emission 
lines have ever been observed of the ``microquasars'' that are known 
to occasionally produce relativistic bi-polar jets with superluminal 
motion, if the observed pair of emission lines might indicate the 
formation of bi-polar outflows in 4U 1630-47 during the peak of its 
1996 outburst. Insufficient coverage at X-rays is unlikely the answer, 
since at least in the case of GRS 1915+105 the source was regularly
monitored by RXTE during a period when superluminal radio jets were 
detected (Fender et al. 1999). In fact, the observations of GRS
1915+105 at other times provided evidence for more subtle mass 
ejection events, combined with observations at other wavelengths 
(Eikenberry et al. 1998; Mirabel et al. 1998), yet no Doppler-shifted 
X-ray lines were reported. Either such lines were overlooked in the
published work based on these observations, or they were simply not
present. We think that the latter is more likely, especially 
considering the fact that no reliable detection has been reported of
any emission lines in microquasars. It has been speculated that the 
general lack (or the
weakness) of iron $K_{\alpha}$ lines in BHCs can be attributed to
the high ionization state of matter in the vicinity of central black
holes due to relatively high X-ray luminosity (Ross \& Fabian 1993; 
Matt, Fabian, \& Ross 1996). It is known that the fluorescent photons 
from Fe XVII--XXIII are very likely to be resonantly absorbed by the 
next ionized species and eventually destroyed by the Auger effect in a
typical environment such as an accretion disk (Ross \& Fabian 1993). 
Consequently, no (or very weak) iron $K_{\alpha}$ lines are expected. 
At even higher ionization state, however, the $K_{\alpha}$ lines from 
Fe XXIV--XXVI can escape rather easily, due to the lack of competing
Auger processes, and thus the yield is quite high. In reality, there 
should exist a range of ionization states in the accretion disk or 
outflows, so the dominating ionization state of matter determines the 
strength of iron $K_{\alpha}$ lines and ultimately whether the lines 
are detectable. It is worth noting that iron absorption lines have 
been detected in microquasars GRS 1915+105 and GRO J1655-40 by ASCA 
(Ebisawa 1996; Ueda et al. 1998), although arguments can be made, in 
the case of GRO J1655-40, for the presence of an emission line, during 
a ``dipping'' period, with a characteristic profile of a disk line 
(see the residual plot for this case in both Ebisawa 1996 and Ueda et 
al. 1998). Much improved spectral capability of X-ray spectrometers 
on future missions, such as {\it Chandra}, {\it XMM}, and {\it Astro-E}, 
can hopefully resolve much of the ambiguity in the interpretation of 
emission lines observed in BHCs.

\acknowledgments
The authors gratefully acknowledge support from NASA through its
Long-Term Space Astrophysics program and RXTE Guest Observer 
program. This work has made use of the results provided by the 
ASM/RXTE teams at MIT and at the RXTE SOF and GOF and of the archival 
databases maintained by the High Energy Astrophysics Science Archive 
Research Center at NASA's Goddard Space Flight Center.

\clearpage
\begin{deluxetable}{lcccc}
\tablecolumns{5}
\tablewidth{0pc}
\tablecaption{PCA Observation Log}
\tablehead{
\colhead{No.}&\colhead{Obs. Id.\tablenotemark{\dag}}&\colhead{Observation Time (UT)}&\colhead{PCUs on}&\colhead{Exposure (s)\tablenotemark{\ddag}}}
\startdata
1 & 02-00 & 05/03/96 20:37:00-22:40:00 & 5 & 2816\tablenotemark{\S} \nl
2 & 03-00 & 05/04/96 12:11:00-14:03:00 & 5 & 3008 \nl
3 & 06-00 & 05/05/96 01:12:00-03:34:00 & 4 & 2368 \nl
4 & 04-00 & 05/06/96 06:59:00-09:39:00 & 4 & 3776 \nl
5 & 05-00 & 05/07/96 16:53:00-18:22:00 & 5 & 3536 \nl
6 & 01-00 & 05/08/96 11:48:00-13:03:00 & 4 & 2992 \nl
7 & 07-00 & 05/09/96 13:38:00-14:48:00 & 4 & 3488 \nl
8 & 08-00 & 05/10/96 16:33:00-17:59:00 & 3 & 3440 \nl
9 & 09-00 & 05/11/96 16:58:00-17:52:00 & 3 & 2784 \nl
10& 11-00 & 05/13/96 05:39:00-08:08:00 & 3 & 4032 \nl
11& 13-00 & 05/15/96 10:10:00-11:28:00 & 3 & 2976 \nl
12& 14-00 & 05/16/96 07:16:00-08:27:00 & 3 & 2880 \nl
13& 15-00 & 05/21/96 02:23:00-03:54:00 & 5 & 2320 \nl
14& 16-00 & 05/29/96 21:03:00-23:10:00 & 5 & 4688 \nl
15& 17-00 & 06/04/96 18:55:00-20:21:00 & 5 & 4064 \nl
16& 18-00 & 08/16/96 08:51:00-09:19:00 & 5 & 1456 \nl
\tablenotetext{\dag}{With prefix 10411-01-}
\tablenotetext{\ddag}{Total amount of on-source time after data filtering and screening}
\tablenotetext{\S}{Excluding the period of dipping activity ($\sim$300 s)}
\enddata
\label{tb:log}
\end{deluxetable}

\begin{deluxetable}{lcccccc}
\tablecolumns{7}
\tablewidth{0pc}
\tablecaption{X-ray Continuum\tablenotemark{\dag}}
\tablehead{
 &  & \multicolumn{2}{c}{Multi-Color Disk\tablenotemark{\ddag}} & \multicolumn{2}{c}{Power Law\tablenotemark{\S}} & \\
\cline{3-4} \cline{5-6} \\
\colhead{No.}&\colhead{$N_H$}&\colhead{$T_{dbb}$}&\colhead{$N_{dbb}$}&\colhead{$\alpha$}&\colhead{$N_{pl}$}&\colhead{$\chi_{\nu}^{2}/dof$} \\
 & $10^{22}\mbox{ }cm^{-2}$ & keV & & & & }
\startdata
1 & $11.5^{+0.6}_{-0.5}$ & $1.26^{+0.01}_{-0.01}$ & $230^{+20}_{-20}$ & $4.06^{+0.11}_{-0.12}$ & $47^{+16}_{-14}$ & 1.37/250 \nl
2 & $10.5^{+0.4}_{-0.3}$ & $1.249^{+0.005}_{-0.005}$ & $220^{+7}_{-8}$ & $3.79^{+0.10}_{-0.10}$ & $15^{+4}_{-4}$ & 0.78/250 \nl
3 & $10.6^{+0.5}_{-0.5}$ & $1.250^{+0.005}_{-0.006}$ & $197^{+6}_{-7}$ & $3.89^{+0.13}_{-0.13}$ & $15^{+4}_{-4}$ & 0.92/200 \nl
4 & $14.1^{+0.5}_{-0.5}$ & $1.31^{+0.01}_{-0.02}$ & $140^{+16}_{-15}$ & $4.77^{+0.08}_{-0.09}$ & $158^{+35}_{-32}$ & 0.89/200 \nl
5 & $13.1^{+0.6}_{-0.6}$ & $1.316^{+0.008}_{-0.009}$ & $175^{+16}_{-18}$ & $4.47^{+0.10}_{-0.11}$ & $97^{+31}_{-24}$ & 0.95/250 \nl
6 & $11.5^{+0.5}_{-0.6}$ & $1.336^{+0.006}_{-0.006}$ & $271^{+13}_{-11}$ & $4.19^{+0.10}_{-0.13}$ & $60^{+16}_{-18}$ & 0.75/200 \nl
7 & $15.1^{+0.6}_{-0.6}$ & $1.40^{+0.03}_{-0.02}$ & $134^{+18}_{-15}$ & $5.04^{+0.09}_{-0.07}$ & $398^{+88}_{-70}$ & 0.75/200 \nl
8 & $15.8^{+0.7}_{-0.6}$ & $1.38^{+0.03}_{-0.02}$ & $133^{+20}_{-20}$ & $4.97^{+0.10}_{-0.09}$ & $346^{+92}_{-75}$ & 1.10/146 \nl
9 & $13.8^{+0.5}_{-0.8}$ & $1.32^{+0.01}_{-0.01}$ & $227^{+18}_{-15}$ & $4.65^{+0.09}_{-0.12}$ & $160^{+43}_{-46}$ & 0.83/146 \nl
10\tablenotemark{\P}& $9.6^{+0.1}_{-0.1}$ & $1.304^{+0.002}_{-0.003}$ & $236^{+2}_{-2}$ & $3.13^{+0.07}_{-0.08}$ & $3.5^{+0.8}_{-0.7}$ & 4.44/146 \nl
11& $13.9^{+0.9}_{-0.6}$ & $1.32^{+0.02}_{-0.01}$ & $178^{+14}_{-23}$ & $4.62^{+0.14}_{-0.09}$ & $134^{+50}_{-29}$ & 0.74/146 \nl
12& $11.9^{+0.6}_{-0.6}$ & $1.319^{+0.006}_{-0.005}$ & $298^{+12}_{-13}$ & $4.18^{+0.11}_{-0.13}$ & $61^{+23}_{-10}$ & 0.72/146 \nl
13& $9.4^{+0.3}_{-0.1}$ & $1.332^{+0.006}_{-0.005}$ & $255^{+5}_{-5}$ & $3.31^{+0.11}_{-0.11}$ & $6^{+2}_{-2}$ & 0.82/250 \nl
14\tablenotemark{\P}& $9.25^{+0.09}_{-0.04}$ & $1.340^{+0.002}_{-0.003}$ & $312^{+2}_{-2}$ & $2.92^{+0.07}_{-0.03}$ & $3.0^{+0.6}_{-0.3}$ & 10.96/250 \nl
15\tablenotemark{\P}&$11.9^{+0.8}_{-0.5}$ & $1.343^{+0.004}_{-0.004}$ & $272^{+7}_{-16}$ & $4.32^{+0.16}_{-0.09}$ & $76^{+42}_{-18}$ & 2.16/250 \nl
16& $7.0^{+1.1}_{-0.9}$ & $0.60^{+0.03}_{-0.02}$ & $765^{+463}_{-267}$ & $1.94^{+0.06}_{-0.06}$ & $0.066^{+0.010}_{-0.008}$ & 0.74/253 \nl 
\tablenotetext{\dag}{The uncertainties are derived by varying parameters 
until $\Delta \chi^2 = 1$. Hence, they correspond to 1$\sigma$ errors.}
\tablenotetext{\ddag}{$T_{dbb}$ is the temperature of the inner edge
of the disk, 
and the normalization $N_{dbb}$ is defined as $(R_{dbb}/D)^2 \cos \theta$,
where $R_{dbb}$ is the radius of the inner edge of the disk in units of km,
$D$ is the distance to the source in units of 10 kpc, and $\theta$ is the 
inclination angle of the disk.}
\tablenotetext{\S}{$\alpha$ is the photon index, and the normalization $N_{pl}$
is defined as photon flux at 1 keV in units of $photons\mbox
{ }cm^{-2}\mbox{ }s^{-1}\mbox{ }keV^{-1}$.}
\tablenotetext{\P}{Note potential large systematic uncertainty in the
measured continuum properties (see text).}
\enddata
\label{tb:cont}
\end{deluxetable}

\begin{deluxetable}{lcccccccc}
\tablecolumns{9}
\tablewidth{0pc}
\tablecaption{Detected Emission Lines\tablenotemark{\dag}}
\tablehead{
 & \multicolumn{3}{c}{Lower-Energy Line} & \multicolumn{3}{c}{Higher-Energy Line} & & \\
\cline{2-4} \cline{4-7} \\
\colhead{No.}&\colhead{$E_l$}&\colhead{$\sigma_l$}&\colhead{$EW_l$}&\colhead{$E_h$}&\colhead{$\sigma_h$}&\colhead{$EW_h$}&\colhead{$E_m$\tablenotemark{\ddag}}&\colhead{$\Delta E$\tablenotemark{\ddag}} \\ 
 &keV&keV&eV&keV&keV&eV&keV&keV }
\startdata
1 & $5.46^{+0.08}_{-0.09}$ & $0.62^{+0.08}_{-0.09}$ & $188^{+42}_{-34}$ & $7.55^{+0.06}_{-0.07}$ & $0.17^{+0.15}_{-0.17}$ & $76^{+14}_{-16}$ & $6.50^{+0.05}_{-0.06}$ & $2.09^{+0.10}_{-0.11}$ \nl
2 & $5.66^{+0.08}_{-0.10}$ & $0.47^{+0.10}_{-0.10}$ & $94^{+22}_{-17}$ & $7.66^{+0.08}_{-0.07}$ & $0.0^{+0.2}_{-0.0}$ & $36^{+8}_{-7}$ & $6.66^{+0.06}_{-0.06}$ & $2.00^{+0.11}_{-0.12}$ \nl
3 & $5.76^{+0.08}_{-0.08}$ & $0.34^{+0.10}_{-0.12}$ & $75^{+17}_{-14}$ & $7.70^{+0.06}_{-0.12}$ & $0.0^{+0.5}_{-0.0}$ & $27^{+12}_{-8}$ & $6.73^{+0.05}_{-0.07}$ & $1.94^{+0.10}_{-0.14}$ \nl
4 & $5.42^{+0.12}_{-0.14}$ & $0.69^{+0.11}_{-0.11}$ & $169^{+53}_{-41}$ & $7.67^{+0.07}_{-0.07}$ & $0.16^{+0.14}_{-0.16}$ & $73^{+18}_{-15}$ & $6.54^{+0.07}_{-0.08}$ & $2.25^{+0.14}_{-0.16}$ \nl
5 & $5.57^{+0.12}_{-0.14}$ & $0.57^{+0.15}_{-0.13}$ & $121^{+26}_{-31}$ & $7.62^{+0.14}_{-0.20}$ & $0.60^{+0.16}_{-0.15}$ & $93^{+43}_{-26}$ & $6.60^{+0.09}_{-0.12}$ & $2.05^{+0.18}_{-0.24}$ \nl
6 & $5.69^{+0.10}_{-0.11}$ & $0.48^{+0.12}_{-0.12}$ & $85^{+24}_{-19}$ & $7.83^{+0.08}_{-0.09}$ & $0.0^{+2.2}_{-0.0}$ & $35^{+11}_{-9}$ & $6.76^{+0.06}_{-0.07}$ & $2.14^{+0.13}_{-0.14}$ \nl
7 & $5.21^{+0.15}_{-0.19}$ & $0.84^{+0.14}_{-0.12}$ & $273^{+92}_{-66}$ & $7.64^{+0.06}_{-0.08}$ & $0.38^{+0.10}_{-0.08}$ & $140^{+19}_{-27}$ & $6.42^{+0.08}_{-0.10}$ & $2.43^{+0.16}_{-0.21}$ \nl
8 & $5.46^{+0.14}_{-0.17}$ & $0.65^{+0.16}_{-0.13}$ & $162^{+70}_{-47}$ & $7.63^{+0.09}_{-0.06}$ & $0.46^{+0.13}_{-0.17}$ & $131^{+50}_{-32}$ & $6.54^{+0.08}_{-0.09}$ & $2.17^{+0.17}_{-0.18}$ \nl
9 & $5.75^{+0.11}_{-0.12}$ & $0.36^{+0.14}_{-0.18}$ & $64^{+22}_{-17}$ & $7.73^{+0.08}_{-0.09}$ & $0.09^{+0.27}_{-0.09}$ & $46^{+19}_{-13}$ & $6.74^{+0.07}_{-0.08}$ & $1.98^{+0.14}_{-0.15}$ \nl
10\tablenotemark{\S}& $5.79^{+0.02}_{-0.03}$ & $0.31^{+0.04}_{-0.04}$ & $51^{+4}_{-4}$ & $7.80^{+0.08}_{-0.07}$ & $0.0^{+0.1}_{-0.1}$ & $12^{+4}_{-2}$ & $6.80^{+0.04}_{-0.04}$ & $2.01^{+0.08}_{-0.08}$ \nl
11& $5.63^{+0.11}_{-0.15}$ & $0.48^{+0.16}_{-0.13}$ & $89^{+41}_{-24}$ & $7.65^{+0.08}_{-0.10}$ & $0.18^{+0.24}_{-0.18}$ & $57^{+30}_{-15}$ & $6.64^{+0.07}_{-0.09}$ & $2.02^{+0.14}_{-0.18}$ \nl
12& $5.78^{+0.10}_{-0.10}$ & $0.25^{+0.14}_{-0.25}$ & $55^{+18}_{-13}$ & $7.76^{+0.11}_{-0.10}$ & $0.0^{+0.3}_{-0.0}$ & $29^{+10}_{-9}$ & $6.77^{+0.07}_{-0.07}$ & $1.98^{+0.15}_{-0.14}$ \nl
13& $5.78^{+0.08}_{-0.10}$ & $0.21^{+0.13}_{-0.21}$ & $50^{+12}_{-10}$ & $7.85^{+0.22}_{-0.19}$& $0.0^{+0.3}_{-0.0}$& $10^{+7}_{-6}$ & $6.82^{+0.12}_{-0.11}$ & $2.07^{+0.23}_{-0.21}$ \nl
14\tablenotemark{\S}& $5.82^{+0.02}_{-0.01}$ & $0.30^{+0.03}_{-0.03}$ & $54^{+3}_{-3}$ & $8.26^{+0.07}_{-0.11}$ & $0.66^{+0.12}_{-0.11}$ & $29^{+11}_{-4}$ & $7.04^{+0.04}_{-0.06}$ & $2.44^{+0.07}_{-0.11}$ \nl
15& $5.79^{+0.08}_{-0.03}$ & $0.25^{+0.13}_{-0.17}$ & $50^{+17}_{-10}$ & $7.95^{+0.12}_{-0.19}$ & $0.5^{+0.3}_{-0.2}$ & $46^{+19}_{-15}$ & $6.87^{+0.07}_{-0.10}$ & $2.16^{+0.14}_{-0.19}$ \nl
16& $6.50^{+0.06}_{-0.08}$ & $0.58^{+0.16}_{-0.11}$ & $453^{+131}_{-79}$ & -- & -- & -- & $6.50^{+0.06}_{-0.08}$ & -- \nl
\tablenotetext{\dag}{Listed properties include line energy, width (FWHM), and 
equivalent width (EW). The uncertainties correspond to 1$\sigma$
errors; see notes for Table~\ref{tb:cont}.}
\tablenotetext{\ddag}{$E_m = (E_l + E_h)/2$, $\Delta E = E_h - E_l$}
\tablenotetext{\S}{Note potential large systematic uncertainty in the
measured line properties (see text).}
\enddata
\label{tb:line}
\end{deluxetable}

\begin{deluxetable}{lcccccccc}
\tablecolumns{9}
\tablewidth{0pc}
\tablecaption{Average "Dip" Spectrum\tablenotemark{\dag}}
\tablehead{
 & \multicolumn{2}{c}{Multi-Color Disk} & \multicolumn{2}{c}{Power Law} & \multicolumn{3}{c}{Emission Line} & \\
\cline{2-3} \cline{4-5} \cline{6-8} \\
\colhead{$N_H$}&\colhead{$T_{dbb}$}&\colhead{$N_{dbb}$}&\colhead{$\alpha$}&\colhead{$N_{pl}$}&\colhead{$E$}&\colhead{$\sigma$}&\colhead{$EW$}&\colhead{$\chi_{\nu}^{2}/dof$} \\
$10^{22}\mbox{ }cm^{-2}$ & keV & & & &keV&keV&eV&}
\startdata
$10.9^{+0.9}_{-1.2}$&$1.50^{+0.27}_{-0.18}$&$18^{+33}_{-12}$&$4.21^{+0.19}_{-0.23}$&$48^{+23}_{-21}$&$5.15^{+0.29}_{-0.33}$&$1.1^{+0.1}_{-0.2}$&$620^{+383}_{-243}$ & 0.84/253
\tablenotetext{\dag}{See tables~\ref{tb:cont} and \ref{tb:line} for the 
definitions and units of the listed quantities. The uncertainties
again correspond to 1$\sigma$ errors.} 
\enddata
\label{tb:dip}
\end{deluxetable}

\clearpage

\end{document}